\documentclass[journal]{IEEEtran}
\usepackage{cite}

\ifCLASSINFOpdf
  \usepackage[pdftex]{graphicx}
  \DeclareGraphicsExtensions{.pdf,.jpeg,.png}
\else
  \usepackage[dvips]{graphicx}
  \DeclareGraphicsExtensions{.eps}
\fi
\usepackage{amsmath}
\begin{document}
%
\title{Finite Element Modeling of Metasurfaces with Generalized Sheet Transition Conditions}
%
%
%

\author{Srikumar~Sandeep,~\IEEEmembership{Member,~IEEE,}
        Jian~Ming Jin,~\IEEEmembership{Fellow,~IEEE,}
        and~Christophe~Caloz,~\IEEEmembership{Fellow,~IEEE}
\thanks{S.\ Sandeep and C.\ Caloz (e-mail: christophe.caloz@polymtl.ca) are with Polytechnique, Montreal.}%
\thanks{J.-M.\ Jin is with the Center for Computational Electromagnetics, Department of Electrical and Computer Engineering, University of Illinois at Urbana-Champaign, Urbana, IL 61801-2991, USA (e-mail: j-jin1@illinois.edu).}}%

%
%

\markboth{IEEE TRANSACTIONS ON ANTENNAS AND PROPAGATION}
{Shell \MakeLowercase{\textit{et al.}}: Bare Demo of IEEEtran.cls for IEEE Journals}
%



\maketitle

\begin{abstract}
A modeling of metasurfaces in the finite element method (FEM) based on generalized sheet transition conditions (GSTCs) is presented. The discontinuities in electromagnetic fields across a metasurface as represented by the GSTC are modeled by assigning nodes to both sides of the metasurface. The FEM-GSTC formulation in both 1D and 2D domains is derived and implemented. The method is extended to handle more general bianistroptic metasurfaces. The formulations are validated by several illustrative examples. 
\end{abstract}


%
\IEEEpeerreviewmaketitle

\section{Introduction}
%
%
%
%
\IEEEPARstart{M}{etasurfaces} are electrically thin layers with embedded subwavelength-sized scatterers \cite{HollawayMetasurfaceReview,MetasurfaceReview}. They are two-dimensional (2D) reductions of three-dimensional (3D) metamaterials \cite{MetamaterialsSolymar,CalozMTM,3DMTMZedler} and offer much richer funtionalities than traditional frequency selectice surfaces \cite{MunkFSS}. The advantages of metasurfaces over metamaterials include lower loss, lighter weight, and easier fabrication. Applications of metasurfaces include polarization transformers \cite{PfeifferGrbicPolControl}, generalized refraction \cite{NYuLightProp}, broadband absorbers \cite{MetasurfaceBroadbandAbsorber}, spatial waveguides \cite{SpatialWaveguideMetasurface}, remotely-controlled spatial processors \cite{AchouriMetasurfaceEPJ}, aberration free lens \cite{AberrationFreeLens},\cite{AietaAchromaticMS}, flat optical components \cite{FlatOpticsYu}, LED efficiency enhancers \cite{LEDEnhancerMS}, and spatial isolators \cite{SpatialIsolatorMS}. Metasurfaces achieve these functionalities by creating discontinuities in the electromagnetic fields. Such discontinuities can be modeled by generalized sheet transition conditions (GSTCs) \cite{HollawayMetasurfaceReview}, \cite{IdemenBook}. Therefore, it is important to develop numerical modeling of GSTCs for the study of metasurfaces.

The modeling of GSTCs in the finite difference method has been recently reported in \cite{YousefFDFD}. In this paper, we present the modeling of GSTCs in the finite element method (FEM), which is one of the most widely used numerical methods to simulate electromagnetic boundary-value problems (BVP) \cite{JinFEMBook,JinCEMBook}. The FEM is particularly suited for solving practical engineering problems given its ability to model complex geometries with adaptive tetrahedral meshes. At present, commercial electromagnetic simulation softwares can model several boundary conditions, such as perfect electric conductor (PEC), perfect magnetic conductor (PMC), periodic boundary condition (PBC), standard impedance boundary condition (SIBC), radiation boundary condition (RBC), and perfectly matched layer (PML). Recently, a formulation to incorporate a generalized impedance boundary condition (GIBC) has also been proposed \cite{FEMGIBC}. However, no commercial CAD tools have yet incorporated the modeling of GSTCs. Since the FEM is the computational method used in the most frequently used CAD tools and since metasurfaces have become increasingly prominent in electromagnetic engineering, there is clearly a need for the modeling of GSTCs in the FEM framework. It should be noted that physical metasurfaces have a finite subwavelength thickness. Simulating such structures directly would result in very dense meshes around the metasurfaces and hence compromise the simulation efficiency. By replacing a physical metasurface by an equivalent GSTC, the burden of mesh generation can be reduced significantly and the simulation efficiency can be enhanced considerably. This is particularly important in simulation scenarios where multiple metasurfaces are involved or when repetitive simulations are required for optimization \cite{achouri2015general}. 

The organization of the paper is as follows. Section II recalls the GSTC metasurface synthesis equations. This is followed by the FEM-GSTC formulation in 1D and 2D domains in Sections III and IV, respectively. Section V presents some simulation results, and Section VI extends the method to simulate bianisotropic metasurfaces. Conclusions are provided in Section VII. 

 

\section{Metasurface synthesis equations}
For a metasurface placed perpendicular to the  $z$ direction of a cartesian coordinate system,  
GSTCs in their most general form are given as \cite{IdemenBook},\cite{achouri2015general}
\begin{subequations}
\label{GSTCeqns}
\begin{align}
\hat{z} \times \Delta \vec{H} = j\omega \vec{P}_{||} - \hat{z} \times \nabla_{||}M_{z}, \label{GSTC1} \\
\hat{z} \times \Delta \vec{E} = -j\omega \mu \vec{M}_{||} - \hat{z} \times \nabla_{||}\left(\frac{P_{z}}{\epsilon}\right), \label{GSTC2}
\end{align}
\end{subequations}
where $\nabla_{||} = \hat{x} {\partial/ \partial x} + \hat{y}{\partial/ \partial y}$, $\vec{P}$ and $\vec{M}$ are the electric and magnetic polarization densities, respectively, and
$\epsilon$ and $\mu$ are the permittivity and permeability of the surrounding medium. Moreover, $\Delta \vec{E}$ and $\Delta \vec{H}$ represent respectively the differences between $\vec{E}$ and $\vec{H}$ on the two sides of the metasurface. For a problem of reflection and transmission, $\Delta \vec{\psi} = \vec{\psi}^{\mathrm{tr}} - (\vec{\psi}^{\mathrm{ref}} + \vec{\psi}^{\mathrm{inc}})$ with the superscripts ``tr,'' ``ref,'' and ``inc'' denoting the transmitted, reflected, and incident fields, respectively. Throughout this work, the normal components of the polarization densities are assumed to be zero, i.e.\ $P_{z} = M_{z} = 0$ \cite{YousefFDFD},\cite{achouri2015general}. 
The polarization densities can be expressed as 
\begin{subequations} \label{PMEqn}
\begin{equation} \label{Peqn1} 
\vec{P} = \epsilon\overline{\overline{\chi}}_{\mathrm{ee}}\vec{E}_{\mathrm{av}} + \sqrt{\epsilon \mu}\ \overline{\overline{\chi}}_{\mathrm{em}}\vec{H}_{\mathrm{av}}
\end{equation}
\begin{equation} \label{Meqn1}
\vec{M} = \overline{\overline{\chi}}_{\mathrm{mm}}\vec{H}_{\mathrm{av}} + \sqrt{\epsilon \mu}\ \overline{\overline{\chi}}_{\mathrm{me}}\vec{E}_{\mathrm{av}}
\end{equation}
\end{subequations}
where $\overline{\overline{\chi}}_{\mathrm{ee}},\overline{\overline{\chi}}_{\mathrm{mm}},\overline{\overline{\chi}}_{\mathrm{em}}$, and $\overline{\overline{\chi}}_{\mathrm{me}}$ are the electric/magnetic (first e/m subscripts) susceptibility tensors describing the response to the electric/magnetic (second e/m subscripts) excitations, and the subscript ``av'' denotes the average of the fields on both sides of the metasurface, $\vec{\psi}_{\mathrm{av}} = [(\vec{\psi}^{\mathrm{inc}} + \vec{\psi}^{\mathrm{ref}}) + \vec{\psi}^{\mathrm{tr}}]/2$. Substituting (\ref{PMEqn}) into (\ref{GSTCeqns}) results in the following metasurface synthesis equations:
\begin{subequations}
\label{GSTCexpandedeqns}
\begin{equation} \label{GSTCexpandedeqns1}
\begin{split}
\begin{pmatrix}
-\Delta H_{y} \\
\Delta H_{x}
\end{pmatrix}
 = & j\omega\epsilon
\begin{pmatrix}
\chi_{\mathrm{ee}}^{xx} & \chi_{\mathrm{ee}}^{xy} \\
\chi_{\mathrm{ee}}^{yx} & \chi_{\mathrm{ee}}^{yy}
\end{pmatrix}
\begin{pmatrix}
E_{x,\mathrm{av}} \\
E_{y,\mathrm{av}}
\end{pmatrix}
\\ & + j\omega\sqrt{\epsilon \mu}
\begin{pmatrix}
\chi_{\mathrm{em}}^{xx} & \chi_{\mathrm{em}}^{xy} \\
\chi_{\mathrm{em}}^{yx} & \chi_{\mathrm{em}}^{yy}
\end{pmatrix}
\begin{pmatrix}
H_{x,\mathrm{av}} \\
H_{y,\mathrm{av}}
\end{pmatrix} ,
\end{split}
\end{equation} 
\begin{equation}
\label{GSTCexpandedeqns2}
\begin{split}
\begin{pmatrix}
\Delta E_{y} \\
-\Delta E_{x}
\end{pmatrix}
  = & j\omega\mu
\begin{pmatrix}
\chi_{\mathrm{mm}}^{xx} & \chi_{\mathrm{mm}}^{xy} \\
\chi_{\mathrm{mm}}^{yx} & \chi_{\mathrm{mm}}^{yy}
\end{pmatrix}
\begin{pmatrix}
H_{x,\mathrm{av}} \\
H_{y,\mathrm{av}}
\end{pmatrix}
\\ & + j\omega\sqrt{\epsilon \mu}
\begin{pmatrix}
\chi_{\mathrm{me}}^{xx} & \chi_{\mathrm{me}}^{xy} \\
\chi_{\mathrm{me}}^{yx} & \chi_{\mathrm{me}}^{yy}
\end{pmatrix}
\begin{pmatrix}
E_{x,\mathrm{av}} \\
E_{y,\mathrm{av}}
\end{pmatrix}
\end{split}
\end{equation}
\end{subequations}
which are applicable for a general bianisotropic metasurface. In the case of a mono-anisotropic metasurface, $\overline{\overline{\chi}}_{\mathrm{em}} = \overline{\overline{\chi}}_{\mathrm{me}} = 0$. Through out this paper, we have assumed, for simplicity but without loss of generality,  that the cross-polarization terms (i.e. tensor elements with either $xy$ or $yx$ as the superscript) in all the four susceptibility tensors are zero. 

\section{GSTCs in 1D FEM}
\begin{figure}[!h]
\centering
\includegraphics[width=3.5in]{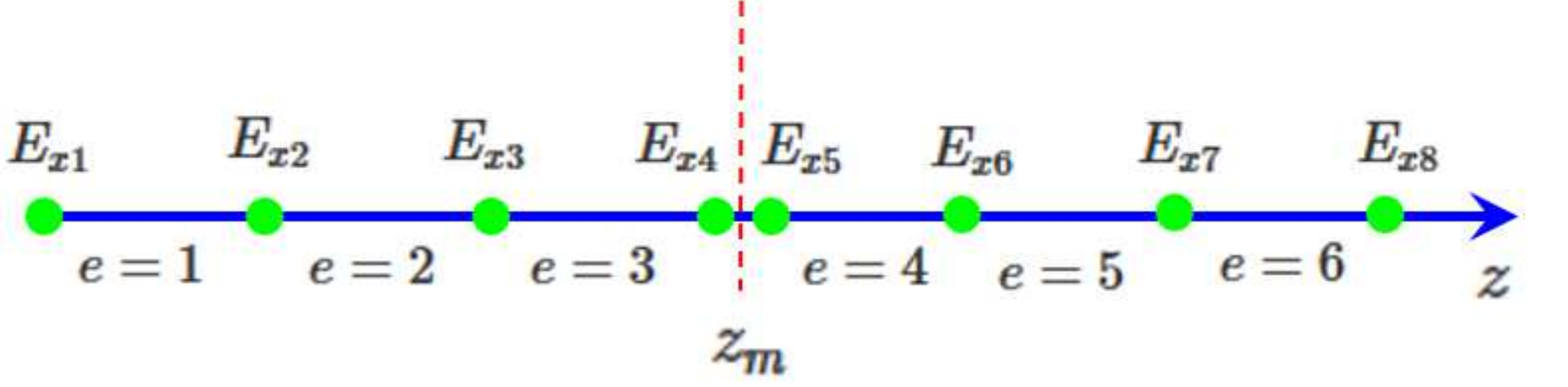}
\caption{1D FEM-GSTC computational domain: $E_{x}$ nodes.}
\label{fig_1DFEM}
\end{figure}

Consider a 1D BVP where $z$ is the only dimension along which material and field variations exist. The computational domain extends from $z=0$ to $z=L$ and a GSTC surface is located at $z=z_{m}$, where $0 < z_{m} < L$. Assuming $E_{x}(z)$ and $H_{y}(z)$ as the field quantities, the scalar wave equation for $E_{x}(z)$ is given by
\begin{equation}
\frac{d}{dz}\left[\frac{1}{\mu_{r}(z)}\frac{dE_{x}}{dz}\right] + k_{o}^{2}\epsilon_{r}(z)E_{x}(z) = 0.
\end{equation}
The FEM domain discretization along with element numbers and global node numbers is shown in Fig.\ \ref{fig_1DFEM}, where $E_{x1}$ is located at $z=0$ and $E_{x8}$ is located at $z=L$. Nodes 4 and 5 are placed on either side of the GSTC surface. 
The electric field in the element $e$ can be expressed by interpolating the nodal electric field values using linear basis functions \cite{JinFEMBook} as
\begin{equation}
E_{x}^{e}(z) = \sum_{j=1}^{2} E_{xj}^{e}N_{j}^{e}(z),
\end{equation}
where $N_{j}^{e}(z)$ is the linear basis function. Using Galerkin's method yields the following linear systems of equations  for the unknown nodal values:  
\begin{subequations}
\label{ExFEMeqns}
\begin{equation}
\begin{bmatrix}
K_{11}^{1} & K_{12}^{1} & 0 & 0 \\
K_{12}^{1} & K_{22}^{1} + K_{11}^{2} & K_{12}^{2} & 0 \\
0 & K_{12}^{2} & K_{22}^{2} + K_{11}^{3} & K_{12}^{3} \\
0 & 0 & K_{12}^{3} & K_{22}^{3}
\end{bmatrix}
\begin{bmatrix}
E_{x1} \\
E_{x2} \\
E_{x3} \\
E_{x4}
\end{bmatrix}
= 
\begin{bmatrix}
b_{1} \\
0 \\
0 \\
b_{4}
\end{bmatrix},
\end{equation}
\begin{equation}
\begin{bmatrix}
K_{11}^{4} & K_{12}^{4} & 0 & 0 \\
K_{12}^{4} & K_{22}^{4} + K_{11}^{5} & K_{12}^{5} & 0 \\
0 & K_{12}^{5} & K_{22}^{5} + K_{11}^{6} & K_{12}^{6} \\
0 & 0 & K_{12}^{6} & K_{22}^{6}
\end{bmatrix}
\begin{bmatrix}
E_{x5} \\
E_{x6} \\
E_{x7} \\
E_{x8}
\end{bmatrix}
= 
\begin{bmatrix}
b_{5} \\
0 \\
0 \\
b_{8}
\end{bmatrix}
.
\end{equation}
\end{subequations}
The matrix elements are calculated by using
\begin{equation}
K^{e}_{ij} = \int_{z^{e}_{1}}^{z^{e}_{2}}\left[\frac{1}{\mu_{r}}\frac{dN^{e}_{i}}{dz}
\frac{dN^{e}_{j}}{dz} - k_{o}^{2}\epsilon_{r}N^{e}_{i}N^{e}_{j}\right] dz,
\label{Keijeqn}
\end{equation}
where $\epsilon_{r}$ and $\mu_{r}$ are the relative permittivity and relative permeability of the medium within the element. The elements of the right-hand-side vectors in (\ref{ExFEMeqns}) are given by
\begin{subequations}
\begin{align}
b_{1} &= -\frac{1}{\mu_{r}}\frac{dE_{x}}{dz}\bigg|_{z=0}, \\
b_{4} &= \frac{1}{\mu_{r}}\frac{dE_{x}}{dz}\bigg|_{z=z_{m}^{-}}, \\
b_{5} &= -\frac{1}{\mu_{r}}\frac{dE_{x}}{dz}\bigg|_{z=z_{m}^{+}}, \\
b_{8} &= \frac{1}{\mu_{r}}\frac{dE_{x}}{dz}\bigg|_{z=L}.
\end{align}
\end{subequations}
The $b_{1}$ and $b_{8}$ can be evaluated by applying the first-order ABC \cite{JinFEMBook}. Such an ABC will excite the computational domain with a plane wave propagating along the $+z$ direction and absorb all the outgoing waves. For the case of an incident wave represented by $e^{-jkz}$, $b_{1}$ and $b_{8}$ are given by
\begin{subequations}
\begin{align}
b_{1} &= \frac{-jk}{\mu_{r}} E_{x1} + \frac{2jk}{\mu_{r}},\\
b_{8} &= \frac{-jk}{\mu_{r}} E_{x8}.
\end{align}
\end{subequations}
The $b_{4}$ and $b_{5}$ can be used to incorporate the metasurface synthesis equations (\ref{GSTCexpandedeqns}). From the Maxwell-Faraday equation, 
\begin{subequations}
\begin{align}
b_{4} &= -j\omega\mu_{o}H_{y}\big|_{z=z_{m}^{-}},\\
b_{5} &= j\omega\mu_{o}H_{y}\big|_{z=z_{m}^{+}}.
\end{align}
\end{subequations}
For a mono-isotropic metasurface, equation (\ref{GSTCexpandedeqns}) reduces to 
\begin{subequations}
\label{MSsynthesisEqn}
\begin{align}
E_{x}|_{z=z_{m}^{+}} - E_{x}|_{z=z_{m}^{-}} &= \frac{-j\omega \mu \chi_{\mathrm{mm}}^{yy}}{2}
\left(H_{y}|_{z=z_{m}^{+}} + H_{y}|_{z=z_{m}^{-}}\right), \\
H_{y}|_{z=z_{m}^{+}} - H_{y}|_{z=z_{m}^{-}} &= \frac{-j\omega \epsilon \chi_{\mathrm{ee}}^{xx}}{2}
\left(E_{x}|_{z=z_{m}^{+}} + E_{x}|_{z=z_{m}^{-}}\right).	
\end{align}
\end{subequations}
Inverting the above equations yields the magnetic field components on either side of the GSTC surface in terms of the electric field components as
\begin{equation}
\begin{bmatrix}
j\omega\mu H_{y}|_{z_{m}^{+}} \\
-j\omega\mu H_{y}|_{z_{m}^{-}} 
\end{bmatrix}
=
\begin{bmatrix}
A & B \\
B & A
\end{bmatrix}
\begin{bmatrix}
E_{x}|_{z_{m}^{+}} \\
E_{x}|_{z_{m}^{-}}
\end{bmatrix}
\end{equation}
with $A$ and $B$ given by 
\begin{subequations}
\begin{equation}
A = \frac{k^{2}\chi_{\mathrm{ee}}^{xx}\chi_{\mathrm{mm}}^{yy}-4}{4\chi_{\mathrm{mm}}^{yy}},
\end{equation}
\begin{equation}
B = \frac{k^{2}\chi_{\mathrm{ee}}^{xx}\chi_{\mathrm{mm}}^{yy}+4}{4\chi_{\mathrm{mm}}^{yy}}.
\end{equation}
\end{subequations}
With these, $b_{4}$ and $b_{5}$ can be written in terms of the electric field values adjacent to the GSTC surface as 
\begin{subequations}
\label{1DGSTCEqn}
\begin{align}
b_{4} = \frac{B}{\mu_{r}} E_{x5} + \frac{A}{\mu_{r}} E_{x4}, \\
b_{5} = \frac{A}{\mu_{r}} E_{x5} + \frac{B}{\mu_{r}} E_{x4}. 
\end{align}
\end{subequations}
The above two equations incorporate GSTC in 1D FEM. In the code, the locations of the nodes corresponding to $E_{x4}$ and $E_{x5}$ can be at $z = z_{m}$. However, $E_{x4}$ should be used to calculate the field solution in element $3$ and $E_{x5}$ should be used to calculate the field solution in element $4$.

\section{GSTCs in 2D FEM}
\begin{figure}[!h]
\centering
\includegraphics[width=3.2in]{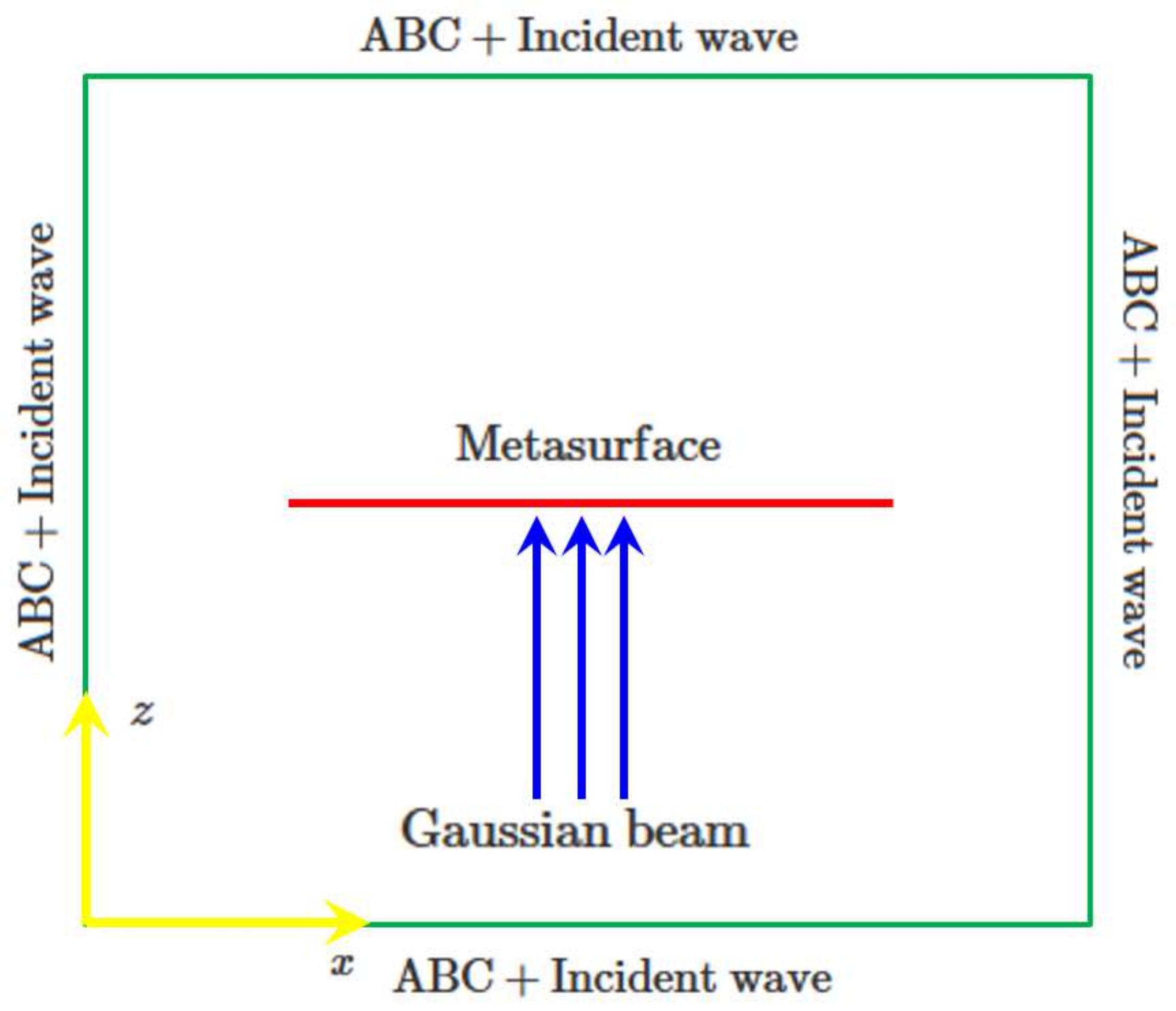}
\caption{2D FEM computational domain.}
\label{fig_2DFEM}
\end{figure}

\begin{figure}[!h]
\centering
\includegraphics[width=3.5in]{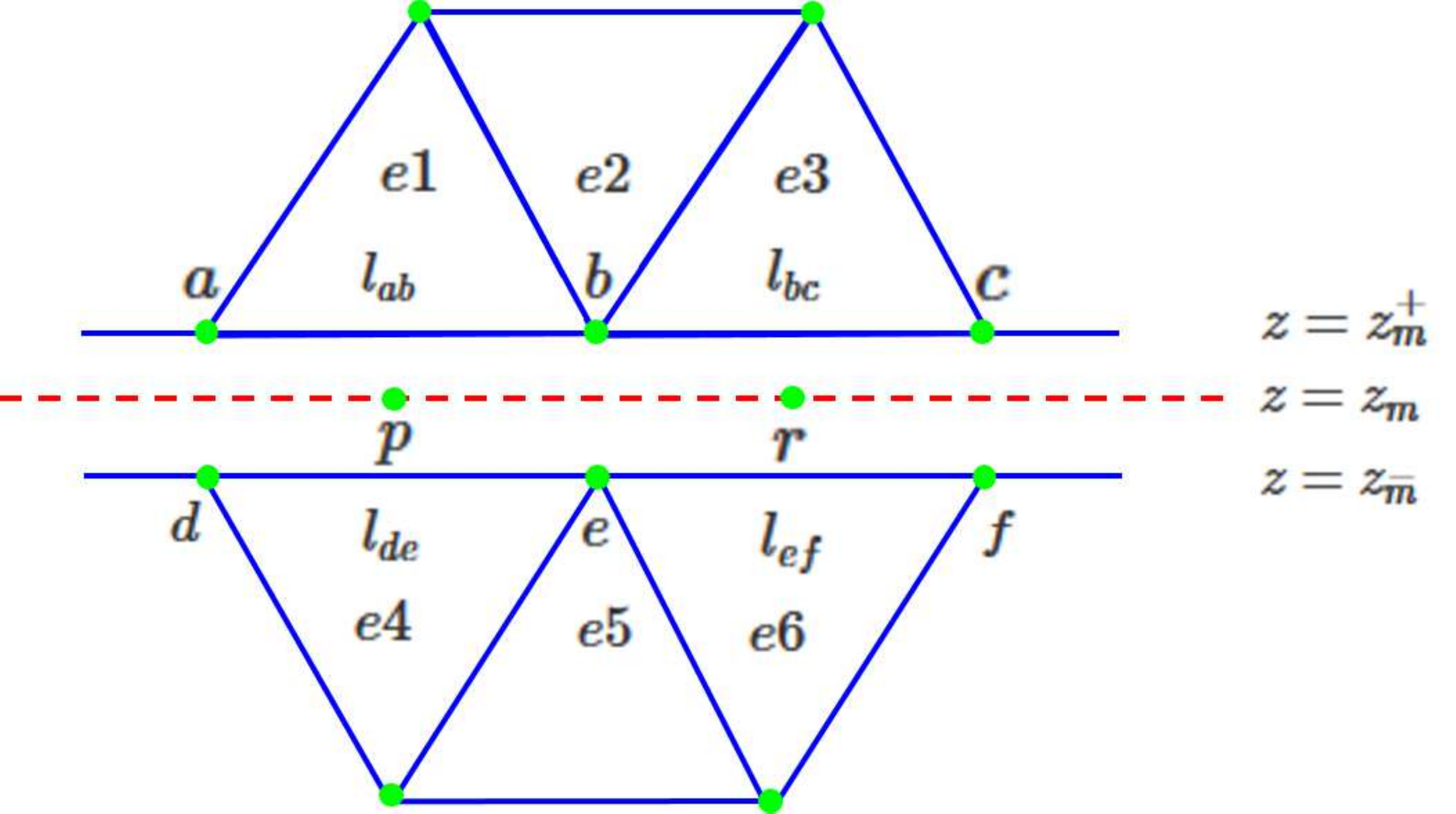}	
\caption{Double nodes around the GSTC surface.}
\label{fig_GSTC2DFEM}
\end{figure}

The 2D computational domain considered in this work is shown in Fig.\ \ref{fig_2DFEM}, where a finite-sized GSTC surface is located at $z=z_{m}$. 
Assuming a $\mathrm{TE}$ polarization ($E_{x},E_{z},H_{y}$), the wave equation for $H_{y}(x,z)$ reads
\begin{equation}
\label{waveEqn2DFEM}
\begin{split}
\frac{\partial}{\partial x}\left[\frac{1}{\epsilon_{r}(x,z)}\frac{\partial H_{y}}{\partial x}\right] & + \frac{\partial }{\partial z}\left[\frac{1}{\epsilon_{r}(x,z)} \frac{\partial H_{y}}{\partial z}\right] \\
& + k_{o}^{2}\mu_{r}(x,z) H_{y} = 0.
\end{split}
\end{equation}
Applying Galerkin's method to (\ref{waveEqn2DFEM}) results in a system of equations for nodal values of $H_{y}$, which can be written as
\begin{equation}
\begin{bmatrix}
K_{11} & K_{12} & \cdots & K_{1N} \\
K_{21} & K_{22} & \cdots & K_{2N} \\
\vdots & \ddots & \cdots & \vdots \\
K_{N1} & K_{N2} & \cdots & K_{NN}
\end{bmatrix}
\begin{bmatrix}
H_{y1} \\
H_{y2} \\
\vdots \\
H_{yN}
\end{bmatrix}
=
\begin{bmatrix}
g_{1} \\
g_{2} \\
\vdots \\
g_{N}
\end{bmatrix} ,
\end{equation}
where $N$ is the total number of nodes in the computational domain. The calculation of the elements of the stiffness matrix, $K_{ij}$, is same as that of 2D FEM without any GSTC surfaces \cite{JinFEMBook}. The elements $g_{j}$ are non-zero only if the corresponding node (i.e. node $j$) lies on the boundary or if the node lies on a surface (or curve) where there is a field discontinuity \cite{JinFEMBook}. Similar to the 1D case, a GSTC surface is modeled by placing nodes both above and below the GSTC surface as shown in Fig.\ \ref{fig_GSTC2DFEM}. It should be noted that the nodes above and below the GSTC surface do not necessarily need to have different locations. But the nodes above the GSTC surface should be used to calculate the field inside the triangular elements which are located above the GSTC surface and the same applies for the nodes below the GSTC surface.
Consider Fig.\ \ref{fig_GSTC2DFEM}, where the dashed line represents the metasurface. The expression for $g_{b}$ (i.e $g_{j}$ for node $b$) can be written as \cite{JinFEMBook}
\begin{equation}
\label{gb1}
\begin{split}
g_{b} = \int_{0}^{l_{ab}}\left(-\hat{z}\right) \cdot \left[\frac{1}{\epsilon_{r}}\frac{\partial H_{y}}{\partial z} \hat{z}\right] \frac{\xi}{l_{ab}} d\xi  \\
 \quad\quad\quad + \int_{0}^{l_{bc}}\left(-\hat{z}\right) \cdot \left[\frac{1}{\epsilon_{r}}\frac{\partial H_{y}}{\partial z}\hat{z}\right] \left(1-\frac{\xi}{l_{bc}}\right) d\xi
\end{split}
\end{equation}
where the first integral corresponds to edge $ab$ and the second integral corresponds to edge $bc$. The contribution to $g_{b}$ of other edges which are connected to node $b$ vanishes because of field continuity. Further, note that $-\hat{z}$ is the unit normal vector pointing outwards from elements $e1$ and $e3$ for edges $ab$ and $bc$, respectively, ${\xi}/{l_{ab}}$ is the value of 2D linear basis function $N^{e1}_{s}(x,z)$ along edge $ab$, where $s$ is the local node number of global node $b$ in element $e1$, and $1 - {\xi}/{l_{bc}}$ is the value of 2D linear basis function $N^{e3}_{t}(x,z)$ along edge $bc$, where $t$ is the local node number of global node $b$ in element $e3$. 
By using Maxwell's equations, (\ref{gb1}) reduces to
\begin{equation}
\begin{split}
g_{b} = \int_{0}^{l_{ab}} j\omega \epsilon_{o}E_{x}\big|_{z_{m}^{+}}  \frac{\xi}{l_{ab}} d\xi  \\
\quad\quad\quad+ \int_{0}^{l_{bc}}j\omega \epsilon_{o}E_{x}\big|_{z_{m}^{+}}  \left(1-\frac{\xi}{l_{bc}}\right) d\xi,
\label{gb2}
\end{split}
\end{equation} 
where $E_{x}\big|_{z_{m}^{+}}$ is a function of $\xi$. 
The metasurface synthesis equations in (\ref{MSsynthesisEqn}) can be used to obtain expressions for $E_{x}|_{z_{m}^{+}}$ and $E_{x}|_{z_{m}^{-}}$ in  terms of $H_{y}|_{z_{m}^{+}}$ and 
$H_{y}|_{z_{m}^{-}}$ as
\begin{equation}
\begin{bmatrix}
j\omega\epsilon E_{x}|_{z_{m}^{+}} \\
-j\omega\epsilon E_{x}|_{z_{m}^{-}} 
\end{bmatrix}
=
\begin{bmatrix}
C & D \\
D & C
\end{bmatrix}
\begin{bmatrix}
H_{y}|_{z_{m}^{+}} \\
H_{y}|_{z_{m}^{-}}
\end{bmatrix}
\end{equation}
where the coefficients $C$ and $D$ are given by 
\begin{subequations}
\begin{equation}
C = \frac{k^{2}\chi_{\mathrm{ee}}^{xx}\chi_{\mathrm{mm}}^{yy}-4}{4\chi_{\mathrm{ee}}^{xx}},
\end{equation}
\begin{equation}
D = \frac{k^{2}\chi_{\mathrm{ee}}^{xx}\chi_{\mathrm{mm}}^{yy}+4}{4\chi_{\mathrm{ee}}^{xx}}.
\end{equation}
\end{subequations}
In the segment $ab$, $E_{x}\big|_{z_{m}^{+}}$ will be denoted as $E_{x}^{ab}(\xi)$ and in the segment $bc$, $E_{x}\big|_{z_{m}^{+}}$ will be denoted as $E_{x}^{bc}(\xi)$. Following this notation, we obtain
\begin{subequations}
\label{Eyeqns}
\begin{align}
j\omega\epsilon_{o} E_{x}^{ab}(\xi) = \frac{C_{p}}{\epsilon_{r}}H_{y}^{ab}(\xi)
+ \frac{D_{p}}{\epsilon_{r}}H_{y}^{de}(\xi) \label{Exabeqn}\\
j\omega\epsilon_{o} E_{x}^{bc}(\xi) = \frac{C_{r}}{\epsilon_{r}}H_{y}^{bc}(\xi)
+ \frac{D_{r}}{\epsilon_{r}}H_{y}^{ef}(\xi) \label{Exbceqn}
\end{align}
\end{subequations}
where it has been assumed that the edges $ab$ and $bc$ are short enough such that the coefficients $C$ and $D$ are approximately constant on these segments. Therefore, $C_{p},D_{p},C_{r},$ and $D_{r}$ are the values of the coefficients $C$ and $D$ at the edge midpoints $p$ and $r$, respectively, as shown in Fig.\ 3. Since linear finite elements are used, linear interpolation can be used to obtain the values of $H_{y}$ at the edges:
\begin{subequations}
\label{Hyeqns}
\begin{align}
H_{y}^{ab}(\xi) &= H_{ya}\left(1-\frac{\xi}{l_{ab}}\right) + H_{yb}\frac{\xi}{l_{ab}}\ , \ 0 \le \xi \le l_{ab}, \label{Hyabeqn}\\
H_{y}^{de}(\xi) &= H_{yd}\left(1-\frac{\xi}{l_{ab}}\right) + H_{ye}\frac{\xi}{l_{ab}}\ , \ 0 \le \xi \le l_{ab}, \label{Hydeeqn}\\
H_{y}^{bc}(\xi) &= H_{yb}\left(1-\frac{\xi}{l_{bc}}\right) + H_{yc}\frac{\xi}{l_{bc}}\ , \ 0 \le \xi \le l_{bc}, \label{Hybceqn}\\
H_{y}^{ef}(\xi) &= H_{ye}\left(1-\frac{\xi}{l_{bc}}\right) + H_{yf}\frac{\xi}{l_{bc}}\ , \ 0 \le \xi \le l_{bc}. \label{Hyefeqn}
\end{align}
\end{subequations}
In the above equations, it is assumed that $l_{ab} = l_{de}$ and $l_{bc} = l_{ef}$, which is equivalent of having an identical mesh just above and below the GSTC surface. Substituting (\ref{Hyeqns}) into (\ref{Eyeqns}) and followed by further substitution into the integrals in (\ref{gb2}), an expression for $g_{b}$ can be obtained as
\begin{equation}
\begin{split}
g_{b} = \frac{C_{p}l_{ab}}{6\epsilon_{r}}H_{ya} + \frac{D_{p}l_{ab}}{6\epsilon_{r}}H_{yd} + 
\left(\frac{C_{p}l_{ab}}{3\epsilon_{r}} + \frac{C_{r}l_{bc}}{3\epsilon_{r}}\right)H_{yb}  \\
+ \left(\frac{D_{p}l_{ab}}{3\epsilon_{r}} + \frac{D_{r}l_{bc}}{3\epsilon_{r}}\right)H_{ye} + 
\frac{C_{r}l_{bc}}{6\epsilon_{r}}H_{yc} + \frac{D_{r}l_{bc}}{6\epsilon_{r}}H_{yf}.
\end{split}
\end{equation}
By following the same procedure, an expression for $g_{e}$ can be obtained as
\begin{equation}
\begin{split}
g_{e} = \frac{C_{p}l_{de}}{6\epsilon_{r}}H_{yd} + \frac{D_{p}l_{de}}{6\epsilon_{r}}H_{ya} + 
\left(\frac{C_{p}l_{de}}{3\epsilon_{r}} + \frac{C_{r}l_{ef}}{3\epsilon_{r}}\right)H_{ye} \\
+ \left(\frac{D_{p}l_{ab}}{3\epsilon_{r}} + \frac{D_{r}l_{bc}}{3\epsilon_{r}}\right)H_{yb} + 
\frac{C_{r}l_{ef}}{6\epsilon_{r}}H_{yf} + \frac{D_{r}l_{ef}}{6\epsilon_{r}}H_{yc}.
\end{split}
\end{equation}
The evaluation of $g_{b}$ and $g_{e}$ for every node above and below the GSTC surface completes the incorporation of GSTC into 2D FEM.

\section{Simulation Results}
This section presents simulation examples to validate FEM-GSTC in both 1D and 2D domains. The metasurface susceptibilities are synthesized by using (\ref{MSsynthesisEqn}). For vacuum on either side of the metasurface, the susceptibilites are obtained as
\begin{subequations}
\label{MetasurfaceSynthExample}	
\begin{equation}
\chi_{\mathrm{ee}}^{xx} = \frac{2}{j\omega\epsilon_{o}}\left[\frac{H_{y}^{\mathrm{inc}} + H_{y}^{\mathrm{ref}} - H_{y}^{\mathrm{tr}}}{E_{x}^{\mathrm{inc}} + E_{x}^{\mathrm{ref}} + E_{x}^{\mathrm{tr}}}\right],
\end{equation}
\begin{equation}
\chi_{\mathrm{mm}}^{yy} = \frac{2}{j\omega\mu_{o}}\left[\frac{E_{x}^{\mathrm{inc}} + E_{x}^{\mathrm{ref}} - E_{x}^{\mathrm{tr}}}{H_{y}^{\mathrm{inc}} + H_{y}^{\mathrm{ref}} + H_{y}^{\mathrm{tr}}}\right].
\end{equation}
\end{subequations}
The simulation frequency is 5 GHz. For both 1D and 2D codes, a simple first-order analytical ABC is used, although the use of the second-order  ABC or PML would  result in more accurate results. 

\subsection{1D Example}
\begin{figure}[!h]
\centering
\includegraphics[width=3.7in]{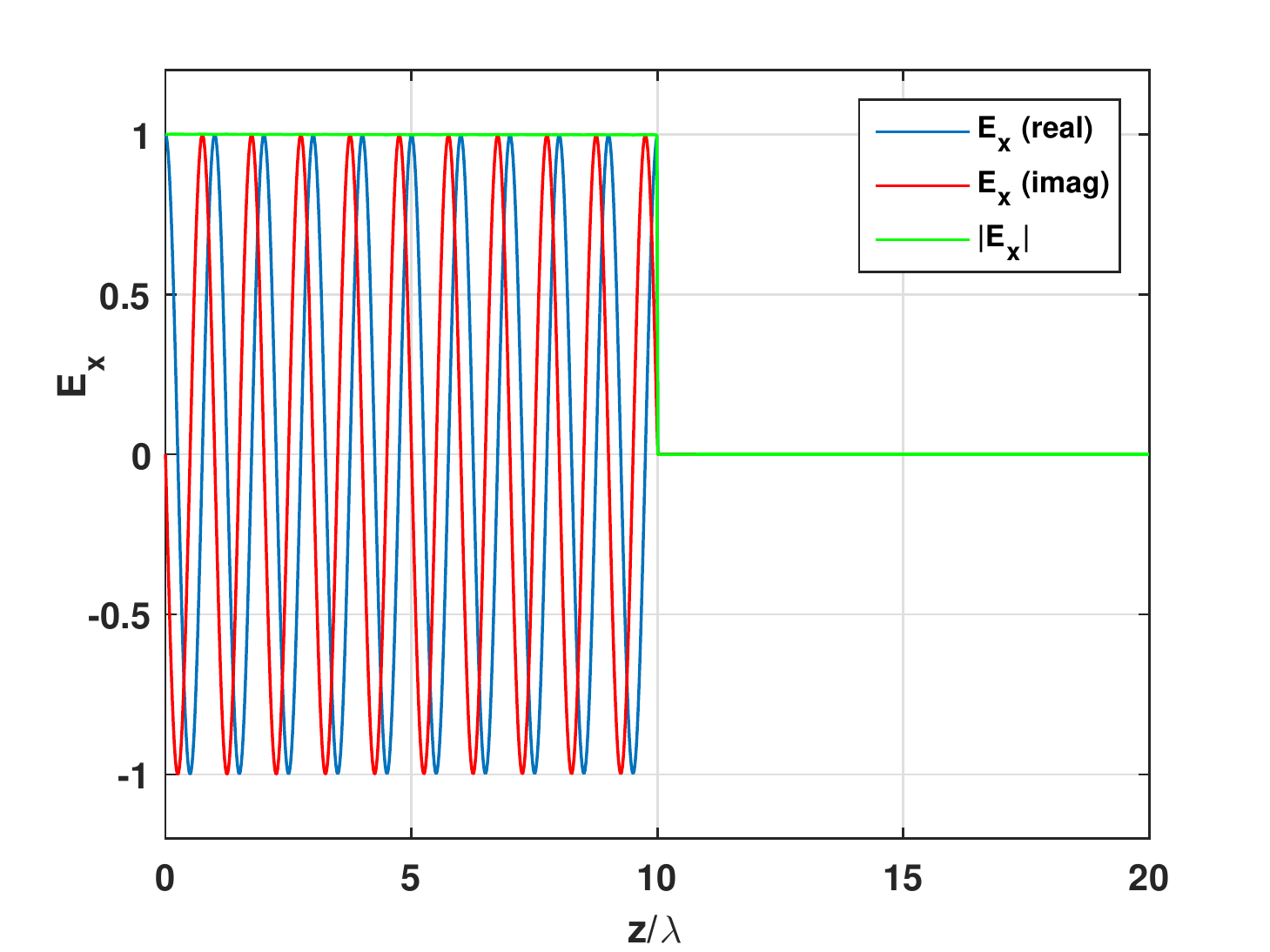}
\caption{1D FEM-GSTC: Fully absorbing metasurface.}
\label{fig_fullAbs1D}
\end{figure}

For validating 1D FEM-GSTC, a fully absorbing metasurface is simulated, for which the reflected and transmitted fields are zero in (\ref{MetasurfaceSynthExample}). The susceptibilities for such a metasurface are thus found as $\chi_{\mathrm{ee}}^{xx} = \chi_{\mathrm{mm}}^{yy} = {2}/{jk_{o}}$, whose negative imaginary nature indicate dissipation \footnote{This may be easily verified by setting $\epsilon_\text{r}=1+\chi_\text{ee,im}^{xx}$. We have then $k_z=nk_0=\sqrt{\epsilon_\text{r}}k_0=\sqrt{1+j\chi_\text{ee,im}^{xx}}k_0\overset{|\chi_\text{ee,im}^{xx}|\ll1}{=}(1+j\chi_\text{ee,im}^{xx}/2)k_0$. So a negative $\chi_\text{ee,im}^{xx}$ implies a decaying wave along $+z$ direction}. The total length of the computational domain is $20\lambda$. The metasurface is located at $z_{m}=10\lambda$. A plane wave with the electric field of magnitude 1 V/m is incident on the metasurface from the left.  The simulation results are plotted in Fig.\ \ref{fig_fullAbs1D},
where it can be seen that the transmitted field to the right of the metasurface is zero, as specified. On the left side of the metasurface, only the incident wave is present, corresponding to unity-magnitude quadrature real and imaginary phasor parts. If a reflected wave were present, one would observe a partly standing-wave pattern with varying field magnitude.

\subsection{2D Examples}
The dimension of the computational domain used for the two 2D FEM-GSTC examples is $26\lambda \times 26\lambda$. The first-order ABC is used on all of the four boundaries of the rectangular computational domain. A finite-sized GSTC surface is located at $z_{m} = 13\lambda$ with a dimension along the $x$-axis of $20\lambda$. A plane wave multiplied by a Gaussian profile (Gaussian variation along the $x$ direction) is incident on the metasurface from below. 

\begin{figure}[!h]
\centering
\includegraphics[width=3.6in]{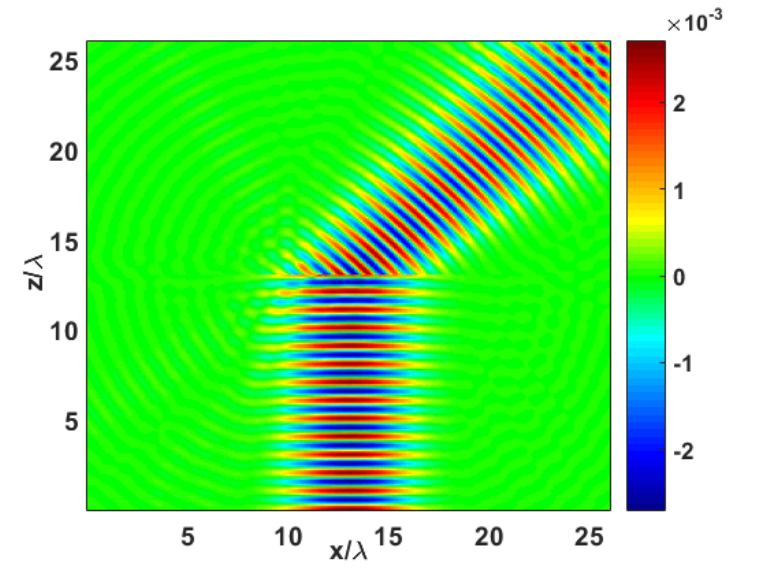}
\caption{Generalized refraction metasurface: Real part of $H_{y}$.}
\label{fig_GenRefReal}
\end{figure}

\begin{figure}[!h]
\centering
\includegraphics[width=3.6in]{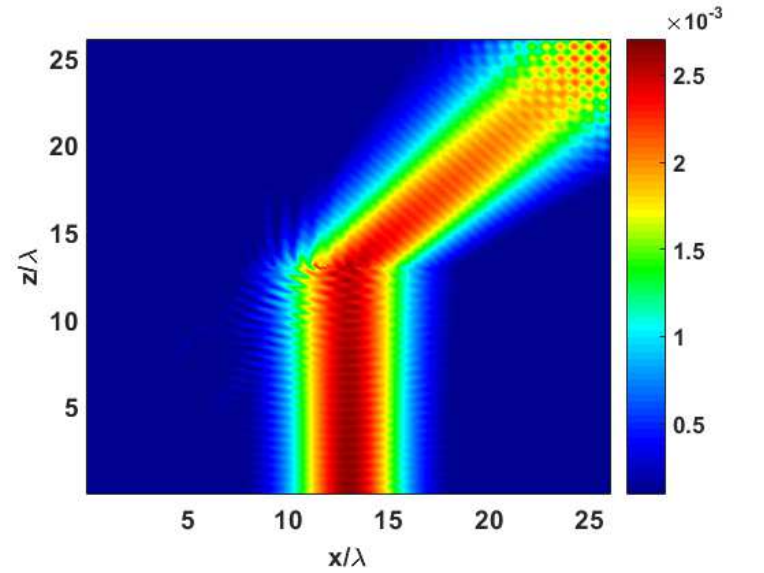}
\caption{Generalized refraction metasurface: Magnitude  of $H_{y}$.}
\label{fig_GenRefMag}
\end{figure}

The first example considers a generalized refracting metasurface with no reflection \cite{achouri2015general}. The metasurface transforms a normally incident plane wave to a plane wave propagating at $\pi/4$ radians to the metasurface normal. The susceptibilities are synthesized using (\ref{MetasurfaceSynthExample}) and explicitly given, plotted and interpreted in \cite{AchouriComparison}. These monoisotropic susceptibilities essentially correspond to a phase-gradient metasurface with loss (and hence negative imaginary susceptibility). The simulation results are plotted in Figs.\ \ref{fig_GenRefReal} and \ref{fig_GenRefMag}.
From these figures, the expected refraction at $\pi/4$ radians is clearly observed. In Fig.\ \ref{fig_GenRefReal}, it can be seen that the metasurface does not create any reflections, as specified. Note that the slight standing-wave pattern observed at the top right corner of Fig.\ 6 is due to the reflections from the absorbing boundary of the computational domain, where the first-order ABC was used. If the second-order ABC or a PML is used, these reflections can be reduced. Since the goal of this work is to implement GSTC in FEM, this issue is not further studied in this case. In Fig.\ \ref{fig_GenRefMag}, weak scattering can be seen at the left end of the Gaussian beam. This could be due to the fact that the susceptibilities were synthesized for the case of plane waves on either side of the metasurface, whereas in the simulation, a plane wave modulated by a transverse Gaussian profile is used. Such a wave is not an exact solution to the vector wave equation. The COMSOL simulation results for a generalized refraction metasurface are reported in \cite{YousefFDFD}. In the COMSOL simulations, the metasurface was represented as a thin slab of a subwavelength thickness. The COMSOL simulation results in \cite{YousefFDFD} showed unspecified refracted beams. Similar to the FDTD-GSTC \cite{YousefFDFD}, the FEM-GSTC does not result in these spurious refracted beams. 

\begin{figure}[!h]
\centering
\includegraphics[width=3.6in]{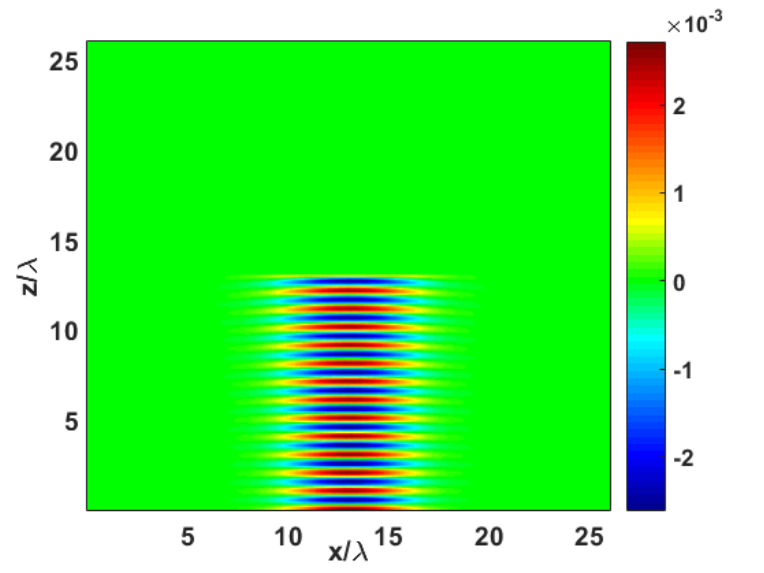}
\caption{Fully absorbing metasurface: Real part of $H_{y}$.}
\label{fig_FullyAbsReal}
\end{figure}

\begin{figure}[!h]
\centering
\includegraphics[width=3.6in]{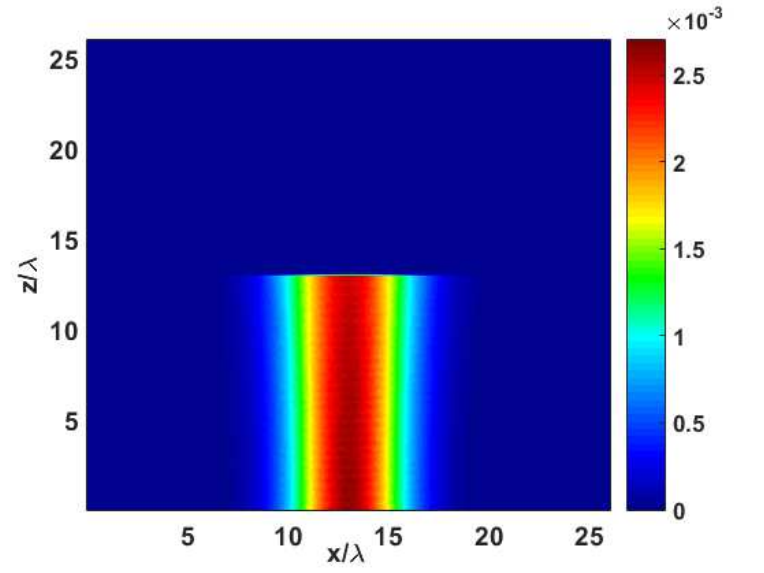}
\caption{Fully absorbing metasurface: Magnitude of $H_{y}$.}
\label{fig_FullyAbsMag}
\end{figure}

The second example is of a fully absorbing metasurface, which is simply the bi-dimensional counterpart of Sec. V.A. The simulation results are shown in Figs.\ \ref{fig_FullyAbsReal} and \ref{fig_FullyAbsMag}.As specified, zero transmission and reflection can be verified in these figures. The COMSOL simulations for the same problem are reported in \cite{YousefFDFD}, where the results showed a partial transmission of the incident beam. 

\section{FEM-GSTC for Bianisotropic Metasurfaces}
The FEM-GSTC described in the previous sections can be extended to model bianisotropic metasurfaces. In this section, we consider bianisotropic metasurfaces where the off-diagonal terms of all the four susceptibility tensors are zero, i.e. $\chi_{\mathrm{ee}}^{xy} = \chi_{\mathrm{ee}}^{yx} = 0$, $\chi_{\mathrm{em}}^{xy} = \chi_{\mathrm{em}}^{yx} = 0$, $\chi_{\mathrm{mm}}^{xy} = \chi_{\mathrm{mm}}^{yx} = 0$, and  $\chi_{\mathrm{me}}^{xy} = \chi_{\mathrm{me}}^{yx} = 0$ in (\ref{GSTCexpandedeqns}). In this case, equation (\ref{GSTCexpandedeqns}) becomes
\begin{subequations}
\label{bianisoeqns}
\begin{equation}
\begin{split}
E_{x}\big|_{z_{m}^{+}} - E_{x}\big|_{z_{m}^{-}} = -\frac{j\omega\mu\chi_{\mathrm{mm}}^{yy}}{2}\left(H_{y}\big|_{z_{m}^{+}} + H_{y}\big|_{z_{m}^{-}}\right) \\
- \frac{j\omega\sqrt{\mu\epsilon}\chi_{\mathrm{me}}^{yy}}{2}\left(E_{y}\big|_{z_{m}^{+}} + E_{y}\big|_{z_{m}^{-}}\right),
\end{split}
\end{equation}
\begin{equation}
\begin{split}
H_{y}\big|_{z_{m}^{+}} - H_{y}\big|_{z_{m}^{-}} = -\frac{j\omega\epsilon\chi_{\mathrm{ee}}^{xx}}{2}\left(E_{x}\big|_{z_{m}^{+}} + E_{x}\big|_{z_{m}^{-}}\right) \\
- \frac{j\omega\sqrt{\mu\epsilon}\chi_{\mathrm{em}}^{xx}}{2}\left(H_{x}\big|_{z_{m}^{+}} + H_{x}\big|_{z_{m}^{-}}\right),
\end{split}
\end{equation}
\begin{equation}
\begin{split}
E_{y}\big|_{z_{m}^{+}} - E_{y}\big|_{z_{m}^{-}} = \frac{j\omega\mu\chi_{\mathrm{mm}}^{xx}}{2}\left(H_{x}\big|_{z_{m}^{+}} + H_{x}\big|_{z_{m}^{-}}\right) \\
+ \frac{j\omega\sqrt{\mu\epsilon}\chi_{\mathrm{me}}^{xx}}{2}\left(E_{x}\big|_{z_{m}^{+}} + E_{x}\big|_{z_{m}^{-}}\right),
\end{split}
\end{equation}
\begin{equation}
\begin{split}
H_{x}\big|_{z_{m}^{+}} - H_{x}\big|_{z_{m}^{-}} = \frac{j\omega\epsilon\chi_{\mathrm{ee}}^{yy}}{2}\left(E_{y}\big|_{z_{m}^{+}} + E_{y}\big|_{z_{m}^{-}}\right) \\
+ \frac{j\omega\sqrt{\mu\epsilon}\chi_{\mathrm{em}}^{yy}}{2}\left(H_{y}\big|_{z_{m}^{+}} + H_{y}\big|_{z_{m}^{-}}\right).
\end{split}
\end{equation}
\end{subequations}
It may be observed from the coupling between field components in these equations that the diagonal elements of the $\overline{\overline{\chi}}_{\mathrm{em}}$ and $\overline{\overline{\chi}}_{\mathrm{me}}$ tensors result in a gyrotropic (chiral) metasurface. 

\begin{figure}[!h]
\centering
\includegraphics[width=3.5in]{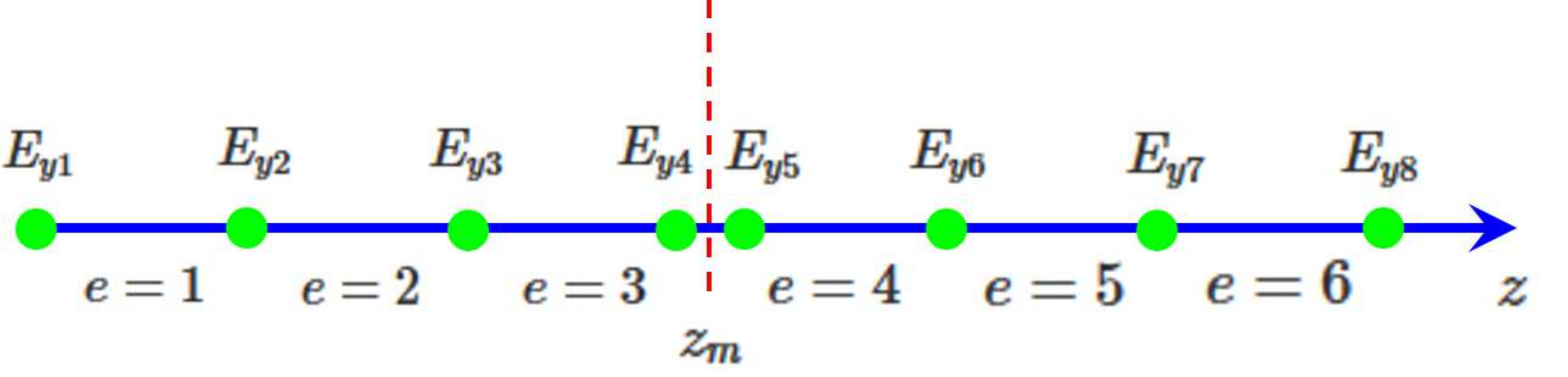}
\caption{1D FEM-GSTC: $E_{y}$ nodes.}
\label{fig_1DFEM_Ey}
\end{figure}

Consider a 1D metasurface problem similar to that in Section III. For a BVP with variations only along the $z$ direction, there are two independent field modes: $\{E_{x},H_{y}\}$ and $\{E_{y},H_{x}\}$. A bianisotropic metasurface induces a  coupling between $\{E_{x},H_{y}\}$ and $\{E_{y},H_{x}\}$ field modes, as seen in (26). Therefore, simulating a bianisotropic metasurface will require a simultaneous processing of the $x$ and $y$ field components. The same principle would apply to a metasurface with off-diagonal components in $\overline{\overline{\chi}}_{\mathrm{ee}}$ and $\overline{\overline{\chi}}_{\mathrm{mm}}$. Thus both $E_{x}$ and $E_{y}$ need to be assigned to the FEM nodes. The domain discretization and node assignment for the case of 8 elements are shown in Figs.\ \ref{fig_1DFEM} and \ref{fig_1DFEM_Ey} for $E_{x}$ and $E_{y}$, respectively.  
The solution vector is $\left[E_{x1}\ \cdots \ E_{x8}\ E_{y1}\ \cdots \ E_{y8}\right]^{T}$. As in Section III, a system of equations can be written for the $E_{y}$ components as
\begin{subequations}
\label{EyFEMEqn}
\begin{equation}
\begin{bmatrix}
K_{11}^{1} & K_{12}^{1} & 0 & 0 \\
K_{12}^{1} & K_{22}^{1} + K_{11}^{2} & K_{12}^{2} & 0 \\
0 & K_{12}^{2} & K_{22}^{2} + K_{11}^{3} & K_{12}^{3} \\
0 & 0 & K_{12}^{3} & K_{22}^{3}
\end{bmatrix}
\begin{bmatrix}
E_{y1} \\
E_{y2} \\
E_{y3} \\
E_{y4}
\end{bmatrix}
= 
\begin{bmatrix}
b_{9} \\
0 \\
0 \\
b_{12}
\end{bmatrix},
\end{equation}
\begin{equation}
\begin{bmatrix}
K_{11}^{4} & K_{12}^{4} & 0 & 0 \\
K_{12}^{4} & K_{22}^{4} + K_{11}^{5} & K_{12}^{5} & 0 \\
0 & K_{12}^{5} & K_{22}^{5} + K_{11}^{6} & K_{12}^{6} \\
0 & 0 & K_{12}^{6} & K_{22}^{6}
\end{bmatrix}
\begin{bmatrix}
E_{y5} \\
E_{y6} \\
E_{y7} \\
E_{y8}
\end{bmatrix}
= 
\begin{bmatrix}
b_{13} \\
0 \\
0 \\
b_{16}
\end{bmatrix}
.
\end{equation}
\end{subequations}
The stiffness matrix elements are given by equation (\ref{Keijeqn}).
The elements of the right-hand-side vectors in (\ref{EyFEMEqn}) are given by
\begin{subequations}
\begin{align}
b_{9} &= -\frac{1}{\mu_{r}}\frac{dE_{y}}{dz}\bigg|_{z=0}, \\
b_{12} &= \frac{1}{\mu_{r}}\frac{dE_{y}}{dz}\bigg|_{z=z_{m}^{-}}, \\
b_{13} &= -\frac{1}{\mu_{r}}\frac{dE_{y}}{dz}\bigg|_{z=z_{m}^{+}}, \\
b_{16} &= \frac{1}{\mu_{r}}\frac{dE_{y}}{dz}\bigg|_{z=L}.
\end{align}
\end{subequations}
Similar to $b_{1}$ and $b_{8}$, the elements $b_{9}$ and $b_{16}$ can be evaluated with the first-order ABC. The other elements $b_{4},b_{5}, b_{12}$, and $b_{13}$ can be evaluated by using Maxwell's equations and GSTCs in (\ref{bianisoeqns}). From the Maxwell-Faraday equation, $b_{4},b_{5}$, $b_{12}$, and $b_{13}$ can be converted to expressions in terms of $H_{x}$ and $H_{y}$ on either side of the metasurface. This is followed by using (\ref{bianisoeqns}) to express $H_{y}|_{z_{m}^{-}}$, $H_{y}|_{z_{m}^{+}}$, $H_{x}|_{z_{m}^{-}}$, and $H_{x}|_{z_{m}^{+}}$ in terms of $E_{y}|_{z_{m}^{-}}$, $E_{y}|_{z_{m}^{+}}$, $E_{x}|_{z_{m}^{-}}$, and $E_{x}|_{z_{m}^{+}}$. The final expressions for $b_{4},b_{5},b_{12}$, and $b_{13}$ are 
\begin{equation}
\label{AmatrixEqn}
\begin{split}
\begin{bmatrix}
b_{4} \\
b_{5} \\
b_{12} \\
b_{13} 
\end{bmatrix}
&= 
\begin{bmatrix}
-j\omega \mu_{o} H_{y}\big|_{z_{m}^{-}} \\
j\omega \mu_{o} H_{y}\big|_{z_{m}^{+}} \\
j\omega \mu_{o} H_{x}\big|_{z_{m}^{-}} \\
-j\omega \mu_{o} H_{x}\big|_{z_{m}^{+}} 
\end{bmatrix}
\\
&= 
\begin{bmatrix}
A_{1} & A_{2} & A_{3} & A_{4} \\
A_{2} & A_{1} & -A_{4} & -A_{3} \\
A_{5} & A_{6} & A_{7} & A_{8} \\
-A_{6} & -A_{5} & A_{8} & A_{7}
\end{bmatrix}
\begin{bmatrix}
E_{x}\big|_{z_{m}^{+}} \\
E_{x}\big|_{z_{m}^{-}} \\
E_{y}\big|_{z_{m}^{+}} \\
E_{y}\big|_{z_{m}^{-}}
\end{bmatrix}
,
\end{split}
\end{equation}
where
\begin{subequations} 
\begin{equation}
A_{1} = \frac{k^{2}\chi_{\mathrm{mm}}^{yy}\left(\chi_{\mathrm{ee}}^{xx}\chi_{\mathrm{mm}}^{xx} - \chi_{\mathrm{em}}^{xx}\chi_{\mathrm{me}}^{xx}\right) + 4\chi_{\mathrm{mm}}^{xx}}{4\mu_{r}\chi_{\mathrm{mm}}^{xx}\chi_{\mathrm{mm}}^{yy}},
\end{equation}
\begin{equation}
A_{2} = \frac{k^{2}\chi_{\mathrm{mm}}^{yy}\left(\chi_{\mathrm{ee}}^{xx}\chi_{\mathrm{mm}}^{xx} - \chi_{\mathrm{em}}^{xx}\chi_{\mathrm{me}}^{xx}\right) - 4\chi_{\mathrm{mm}}^{xx}}{4\mu_{r}\chi_{\mathrm{mm}}^{xx}\chi_{\mathrm{mm}}^{yy}},
\end{equation}
\begin{equation}
A_{3} = -\frac{jk_{o}}{2}\sqrt{\frac{\epsilon_{r}}{\mu_{r}}}\left(\frac{\chi_{\mathrm{em}}^{xx}\chi_{\mathrm{mm}}^{yy} - \chi_{\mathrm{mm}}^{xx}\chi_{\mathrm{me}}^{yy}}{\chi_{\mathrm{mm}}^{xx}\chi_{\mathrm{mm}}^{yy}}\right),
\end{equation}
\begin{equation}
A_{4} = \frac{jk_{o}}{2}\sqrt{\frac{\epsilon_{r}}{\mu_{r}}}\left(\frac{\chi_{\mathrm{em}}^{xx}\chi_{\mathrm{mm}}^{yy} + \chi_{\mathrm{mm}}^{xx}\chi_{\mathrm{me}}^{yy}}{\chi_{\mathrm{mm}}^{xx}\chi_{\mathrm{mm}}^{yy}}\right),
\end{equation}
\begin{equation}
A_{5} = \frac{jk_{o}}{2}\sqrt{\frac{\epsilon_{r}}{\mu_{r}}}\left(\frac{\chi_{\mathrm{mm}}^{xx}\chi_{\mathrm{em}}^{yy} - \chi_{\mathrm{me}}^{xx}\chi_{\mathrm{mm}}^{yy}}{\chi_{\mathrm{mm}}^{xx}\chi_{\mathrm{mm}}^{yy}}\right),
\end{equation}
\begin{equation}
A_{6} = -\frac{jk_{o}}{2}\sqrt{\frac{\epsilon_{r}}{\mu_{r}}}\left(\frac{\chi_{\mathrm{mm}}^{xx}\chi_{\mathrm{em}}^{yy} + \chi_{\mathrm{me}}^{xx}\chi_{\mathrm{mm}}^{yy}}{\chi_{\mathrm{mm}}^{xx}\chi_{\mathrm{mm}}^{yy}}\right),
\end{equation}
\begin{equation}
A_{7} = -\frac{k^{2}\chi_{\mathrm{mm}}^{xx}\left(\chi_{\mathrm{em}}^{yy}\chi_{\mathrm{me}}^{yy} - \chi_{\mathrm{ee}}^{yy}\chi_{\mathrm{mm}}^{yy}\right) - 4\chi_{\mathrm{mm}}^{yy}}{4\mu_{r}\chi_{\mathrm{mm}}^{xx}\chi_{\mathrm{mm}}^{yy}},
\end{equation}
\begin{equation}
A_{8} = -\frac{k^{2}\chi_{\mathrm{mm}}^{xx}\left(\chi_{\mathrm{em}}^{yy}\chi_{\mathrm{me}}^{yy} - \chi_{\mathrm{ee}}^{yy}\chi_{\mathrm{mm}}^{yy}\right) + 4\chi_{\mathrm{mm}}^{yy}}{4\mu_{r}\chi_{\mathrm{mm}}^{xx}\chi_{\mathrm{mm}}^{yy}}.
\end{equation}
\end{subequations}

\begin{figure}[!h]
\centering
\includegraphics[width=3.5in]{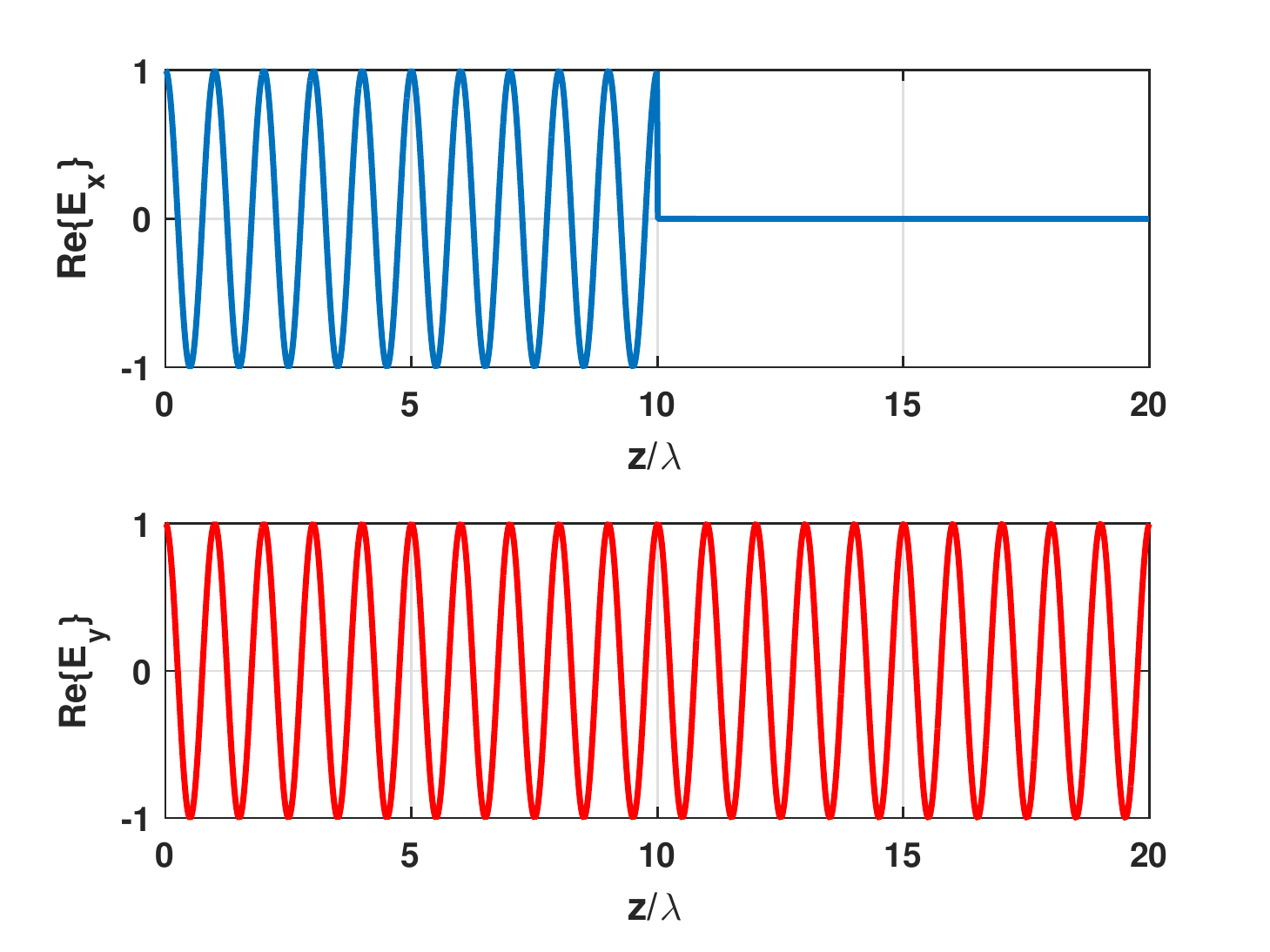}
\caption{FEM-GSTC simulation results for a bianisotropic metasurface.}
\label{fig_bianiso}
\end{figure}

The FEM-GSTC for bianisotropic metasurfaces is validated by a simple example. As seen in (\ref{bianisoeqns}), such a metasurface is characterized by eight free tensorial susceptibility elements for four equations. This corresponds to an underdetermined problem, allowing for the possibility for the metasurface to perform a double transformation, as noted in [21]. We shall consider here a birefringent metasurface \cite{AchouriComparison} that transforms 1) an $x$-polarized incident wave into a $y$-polarized transmitted wave and 2) a $y$-polarized incident wave into a fully absorbed wave, with zero reflection in both cases. In terms of fields, the first transformation reads
\begin{subequations}
\label{wavetr1}
\begin{equation}
E_{x}^{(1)}\big|_{z_{m}^{-}} = e^{-jk_{o}z_{m}} \ , \ H_{y}^{(1)}\big|_{z_{m}^{-}} = \frac{e^{-jk_{o}z_{m}}}{\eta_{o}},
\end{equation}
\begin{equation}
E_{y}^{(1)}\big|_{z_{m}^{-}} = 0 \ , \ H_{x}^{(1)}\big|_{z_{m}^{-}} = 0,
\end{equation}
\begin{equation}
E_{x}^{(1)}\big|_{z_{m}^{+}} = 0 \ , \ H_{y}^{(1)}\big|_{z_{m}^{+}} = 0,
\end{equation}
\begin{equation}
E_{y}^{(1)}\big|_{z_{m}^{+}} = e^{-jk_{o}z_{m}} \ , \ H_{x}^{(1)}\big|_{z_{m}^{+}} = -\frac{e^{-jk_{o}z_{m}}}{\eta_{o}},
\end{equation}
\end{subequations}
and the second transformation reads
\begin{subequations}
\label{wavetr2}
\begin{equation}
E_{x}^{(2)}\big|_{z_{m}^{-}} = 0 \ , \ H_{y}^{(2)}\big|_{z_{m}^{-}} = 0,
\end{equation}
\begin{equation}
E_{y}^{(2)}\big|_{z_{m}^{-}} = e^{-jk_{o}z_{m}} \ , \ H_{x}^{(2)}\big|_{z_{m}^{-}} = -\frac{e^{-jk_{o}z_{m}}}{\eta_{o}},
\end{equation}
\begin{equation}
E_{x}^{(2)}\big|_{z_{m}^{+}} = 0 \ , \ H_{y}^{(2)}\big|_{z_{m}^{+}} = 0,
\end{equation}
\begin{equation}
E_{y}^{(2)}\big|_{z_{m}^{+}} = 0 \ , \ H_{x}^{(2)}\big|_{z_{m}^{+}} = 0.
\end{equation}
\end{subequations}
By substituting (\ref{wavetr1}) and (\ref{wavetr2}) in (\ref{bianisoeqns}), one obtains a linear system of eight equations for the eight unknown susceptibilities, which can be solved to yield
\begin{subequations}
\begin{equation}
\chi_{\mathrm{ee}}^{xx} = \chi_{\mathrm{ee}}^{yy} = -\frac{2j}{k_{o}},
\end{equation}
\begin{equation}
\chi_{\mathrm{mm}}^{xx} = \chi_{\mathrm{mm}}^{yy} = -\frac{2j}{k_{o}},
\end{equation}
\begin{equation}
\chi_{\mathrm{em}}^{xx} = 0 \ , \  \chi_{\mathrm{em}}^{yy} = \frac{4j}{k_{o}},
\end{equation}
\begin{equation}
\chi_{\mathrm{me}}^{xx} = -\frac{4j}{k_{o}} \ , \  \chi_{\mathrm{me}}^{yy} = 0.
\end{equation}
\end{subequations}
The metasurface is illuminated with two plane waves, $E_{x}^{\mathrm{inc}} = e^{-jk_{o}z}$ and $E_{y}^{\mathrm{inc}} = e^{-jk_{o}z}$. The simulation parameters are same as those in Section V. The metasurface is located at $z_{m}=10\lambda$. The real parts of $E_{x}$ and $E_{y}$ are shown in Fig.\ \ref{fig_bianiso},
where it can be observed that on the right side of the metasurface, there is only an $E_{y}$ component. This is due to the fact that the $x$ polarized wave incident on the metasurface from its left side is transformed to a $y$ polarized wave and the $y$ polarized wave incident on the metasurface from its left side is completely absorbed. It can also be seen that for $z < 10\lambda$, there are no reflections.

\section{Conclusion}
In this paper, we presented the FEM modeling of metasurfaces based on GSTCs, where the discontinuities in electromagnetic fields across a metasurface were modeled by assigning nodes to both sides of the metasurface. We derived the FEM-GSTC formulation in both 1D and 2D domains and extended it to handle more general bianistroptic metasurfaces. We also presented several illustrative examples to validate the FEM-GSTC formulation. Future work includes extension of the method to 3D problems and curved metasurfaces. In 3D problems, the susceptibility tensor elements will be functions of both $x$ and $y$. The extension of 2D FEM-GSTC to 3D is straightforward. In 2D, the line integrals along the element edges were used to calculate the right-hand-side of the FEM system of equations. In 3D, surface integrals on the tetrahedral faces should be used to obtain the right-hand-side of the FEM system of equations. Even though this paper assumed zero off-diagonal elements in the susceptibility tensors, the formulation can be easily extended to non-zero off-diagonal elements. In such a case, the matrix elements in (\ref{AmatrixEqn}) will be more involved. It should be noted that in this paper, the GSTC surface is handled by placing field nodes on either side of the GSTC surface. A similar procedure is used in the FDFD formulation in \cite{YousefFDFD}. However, the advantage of FEM-GSTC is its flexibility in placing these nodes conforming to the metasurface geometry. The rectangular Yee cells of FDFD limits this flexibility. Hence FEM-GSTC is more efficient for simulating arbitrarily shaped metasurfaces or in general arbitrarily shaped spatial electromagnetic field discontinuity.


%





\ifCLASSOPTIONcaptionsoff
  \newpage
\fi



\bibliographystyle{IEEEtran}
\bibliography{IEEEabrv,BibDataBase}
%

%

\begin{IEEEbiography}[{\includegraphics[width=1.25in,height=1.75in,clip,keepaspectratio]{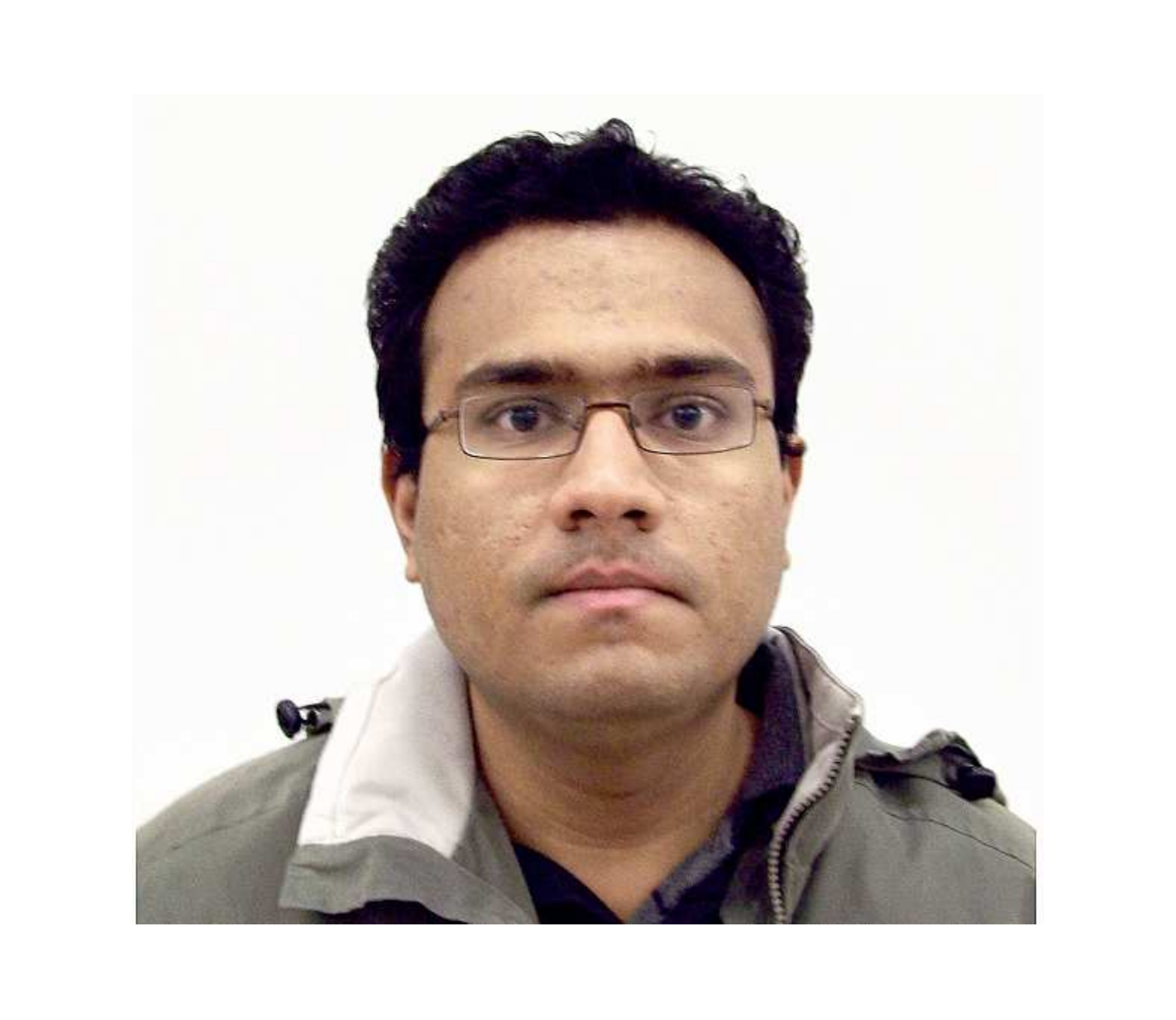}}]{Srikumar Sandeep}
Dr. Srikumar Sandeep received his Ph.D. degree in electrical engineering from University of Colorado, Boulder in 2012. He was a postdoctoral researcher at University of Colorado, Boulder in 2013 and Ecole Polytechnique, Montreal in 2016. He has more than 4 years of industrial experience in software development, embedded systems and signal integrity. He holds 2 US patents. His technical interests include applied and computational electromagnetics. He served as a reviewer for several journals such as \emph{Applied Optics}, \emph{IEEE Antennas and Wireless Propagation Letters}, \emph{Applied Computational Electromagnetics Society (ACES) Journal}, and \textsc{IEEE Transactions on Geoscience and Remote Sensing}

\end{IEEEbiography}

\vfill

\begin{IEEEbiography}
[{\includegraphics[width=1in,height=1.25in,clip,keepaspectratio]{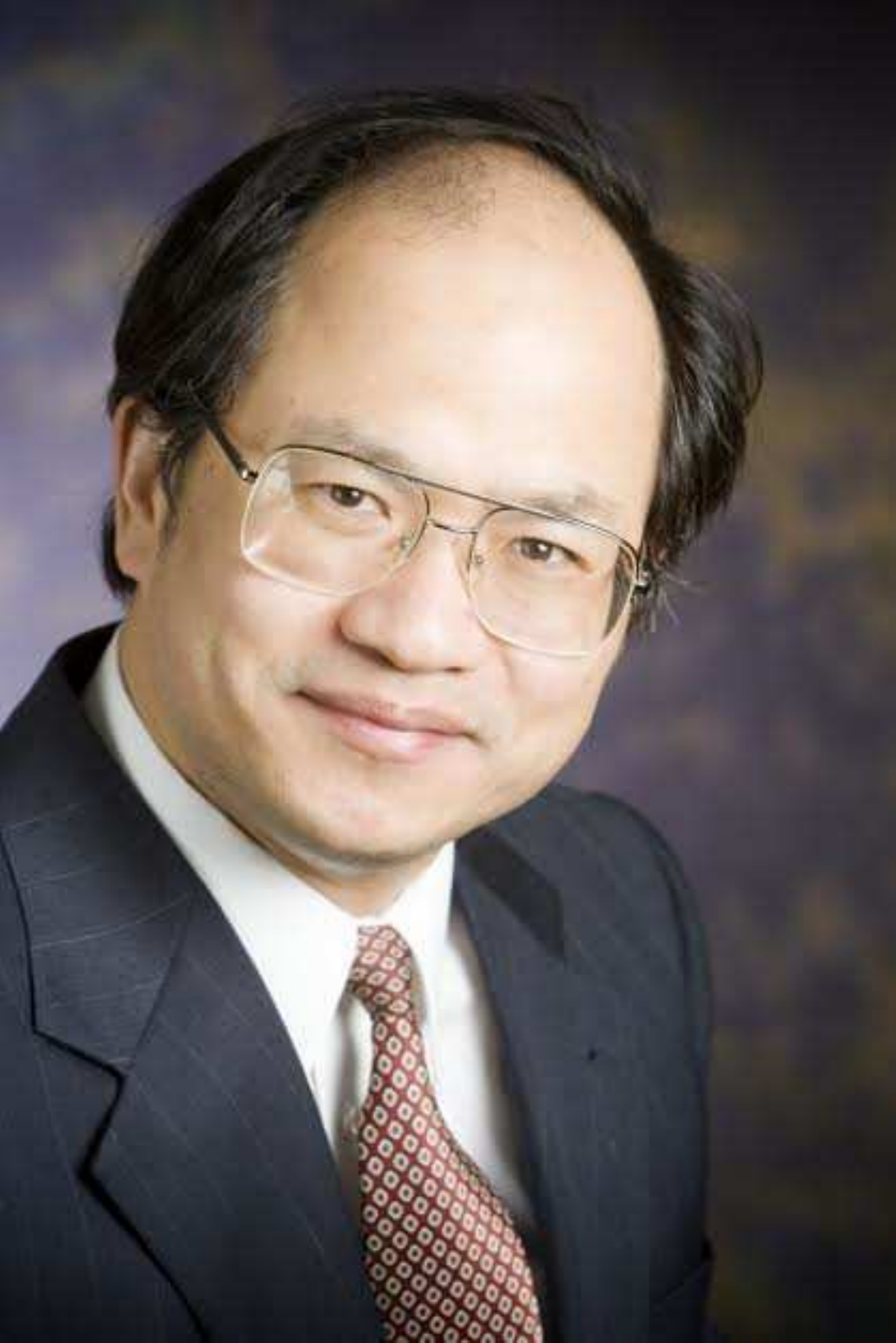}}]{Jian-Ming~Jin}
(S'87-M'89-SM'94-F'01) received his Ph.D. degree in electrical engineering from the University of Michigan, Ann Arbor, in 1989. He joined the University of Illinois at Urbana--Champaign in 1993 and is currently the Y.~T.~Lo Chair Professor of Electrical and Computer Engineering and Director of the Electromagnetics Laboratory and Center for Computational Electromagnetics. He has authored and co-authored over 250 papers in refereed journals and 22 book chapters. He has also authored \emph{The Finite Element Method in Electromagnetics} (Wiley, 1st ed 1993, 2nd ed 2002, 3rd ed 2014), \emph{Electromagnetic Analysis and Design in Magnetic Resonance Imaging} (CRC, 1998), \emph{Theory and Computation of Electromagnetic Fields} (Wiley, 1st ed 2010, 2nd ed 2015), and co-authored \emph{Computation of Special Functions} (Wiley, 1996), \emph{Fast and Efficient Algorithms in Computational Electromagnetics} (Artech, 2001), and \emph{Finite Element Analysis of Antennas and Arrays} (Wiley, 2008). His current research interests include computational electromagnetics, scattering and antenna analysis, electromagnetic compatibility, high-frequency circuit modeling and analysis, bioelectromagnetics, and magnetic resonance imaging. He was elected by ISI as one of the world's most cited authors in 2002.

Dr. Jin is a Fellow of the Electromagnetics Academy and Applied Computational Electromagnetics Society (ACES), and a member of URSI Commission B. He was a recipient of the 1994 National Science Foundation Young Investigator Award, the 1995 Office of Naval Research Young Investigator Award, the 1999 ACES Valued Service Award, the 2014 ACES Technical Achievement Award, the 2016 ACES Computational Electromagnetics Award, the 2015 IEEE Antennas and Propagation Society Chen-To Tai Distinguished Educator Award, and the 2015 IEEE Antennas and Propagation Edward E. Altschuler AP-S Magazine Prize Paper Award. He also received the 1997 Xerox Junior Research Award and the 2000 Xerox Senior Research Award presented by the College of Engineering, University of Illinois at Urbana--Champaign, and was appointed as the first Henry Magnuski Outstanding Young Scholar in the Department of Electrical and Computer Engineering in 1998 and later as a Sony Scholar in 2005. He was appointed as a Distinguished Visiting Professor in the Air Force Research Laboratory in 1999 and was awarded Adjunct, Visiting, Guest, or Chair Professorship by 11 institutions around the world. His name appeared 22 times in the University of Illinois at Urbana--Champaign's List of Excellent Instructors. His students have won the best paper awards in IEEE 16th Topical Meeting on Electrical Performance of Electronic Packaging and 25th, 27th, 31st, and 32nd Annual Review of Progress in Applied Computational Electromagnetics. He served as an Associate Editor and Guest Editor for the \textsc{IEEE Transactions on Antennas and Propagation}, \emph{Radio Science}, \emph{Electromagnetics}, \emph{Microwave and Optical Technology Letters}, and \emph{Medical Physics}.
\end{IEEEbiography}

 \vfill


\begin{IEEEbiography}[{\includegraphics[width=1in,height=1.25in,clip,keepaspectratio]{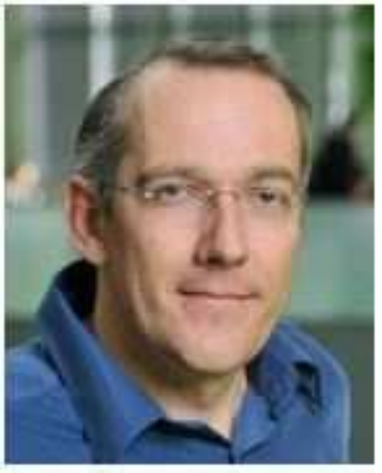}}]{Christophe Caloz}
(F'10) Christophe Caloz received the Diplôme d'Ingénieur en Électricité and the Ph.D. degree from École Polytechnique Fédérale de Lausanne (EPFL), Switzerland, in 1995 and 2000, respectively. From 2001 to 2004, he was a Postdoctoral Research Fellow at the Microwave Electronics Laboratory, University of California at Los Angeles (UCLA). In June 2004, Dr. Caloz joined École Polytechnique of Montréal, where he is now a Full Professor, the holder of a Canada Research Chair (CRC) Tier-I and the head of the Electromagnetics Research Group. He has authored and co-authored over 700 technical conference, letter and journal papers, 13 books and book chapters, and he holds several patents. His works have generated about 20,000 citations, and he has been a Thomson Reuters Highly Cited Researcher. Dr. Caloz was a Member of the Microwave Theory and Techniques Society (MTT-S) Technical Committees MTT-15 (Microwave Field Theory) and MTT-25 (RF Nanotechnology), a Speaker of the MTT-15 Speaker Bureau, the Chair of the Commission D (Electronics and Photonics) of the Canadian Union de Radio Science Internationale (URSI) and an MTT-S representative at the IEEE Nanotechnology Council (NTC). In 2009, he co-founded the company ScisWave (now Tembo Networks). Dr. Caloz received several awards, including the UCLA Chancellor’s Award for Post-doctoral Research in 2004, the MTT-S Outstanding Young Engineer Award in 2007, the E.W.R. Steacie Memorial Fellowship in 2013, the Prix Urgel-Archambault in 2013, the Killam Fellowship in 2016, and many best paper awards with his students at international conferences. He has been an IEEE Fellow since 2010, an IEEE Distinguished Lecturer for the Antennas and Propagation Society (AP-S) since 2014, and a Fellow of the Canadian Academy of Engineering since 2016. He was an Associate Editor of the Transactions on Antennas and Propagation of AP-S in from 2015 to 2017. In 2014, Dr. Caloz was elected as a member of the Administrative Committee of AP-S. His research interests include all fields of theoretical, computational and technological electromagnetics, with strong emphasis on emergent and multidisciplinary topics, including particularly metamaterials, nanoelectromagnetics, exotic antenna systems and real-time radio.
\end{IEEEbiography}


\vfill


\end{document}